\documentclass[review]{elsarticle}

\usepackage{lineno,hyperref}
\modulolinenumbers[5]

\journal{Applied Mathematics and Computation}

\begin{document}

\begin{frontmatter}

\title{Effect of information asymmetry in Cournot duopoly game with bounded rationality}

\author{Masahiko Ueda\corref{mycorrespondingauthor}}
\address{Department of Systems Science, Graduate School of Informatics, Kyoto University, Kyoto 606-8501, Japan}
\ead{ueda.masahiko.5r@kyoto-u.ac.jp}

\begin{abstract}
We investigate the effect of information asymmetry on a dynamic Cournot duopoly game with bounded rationality.
Concretely, we study how one player's possession of information about the other player's behavior in a duopoly affects the stability of the Cournot-Nash equilibrium.
We theoretically and numerically show that the information stabilizes the Cournot-Nash equilibrium and suppresses chaotic behavior in the duopoly.
\end{abstract}

\begin{keyword}
Discrete dynamical systems; Cournot duopoly games; Bounded rationality; Information asymmetry; Complex dynamics
\end{keyword}

\end{frontmatter}


\section{Introduction}
\label{sec:introduction}
An oligopoly is a market where a few firms control the price of a good.
Cournot first introduced the oligopoly model in 1838.
He considered an economy of two firms (players) producing the same good, both firms choosing their respective production to maximize their profit.
Because the profit of each firm depends on its production as well as the production of the other firm, the situation is game theoretic.
Each firm needs to correctly anticipate the behavior of the other firm. 
The equilibrium state between the two players is known as the Cournot-Nash equilibrium; here, each player's response to the other is optimal.
This equilibrium is realized when the two players are sufficiently rational.

However, in reality, the players are not necessarily rational.
Previous studies have examined games in which the players have bounded rationality \cite{Rub1998}.
As a toy model, the Cournot duopoly game model with bounded rationality has recently attracted much attention.
The discrete-time dynamics of a Cournot duopoly game with bounded rationality has been analyzed in Refs. \cite{BisNai2000,BSG1998}.
In this model, two players adjust their outputs step by step in order to increase their respective profits.
Here, the Cournot-Nash equilibrium is not necessarily stable, and even chaotic behavior can occur in some parameter regions, as in dynamic Cournot duopoly games with naive response \cite{Puu1991,Kop1996}.
Following these studies, the model was extended to the case of nonlinear demand function \cite{AHE2002,NaiSbr2006,FGS2015}.
Another extension is the Cournot duopoly game with heterogeneous players, where the two players adopt different decision-making strategies \cite{BisKop2001,AgiEls2003,AgiEls2004,ADN2009,CavNai2014,Els2015}.
In addition, the dynamic Cournot duopoly game with time delay \cite{YasAgi2003}
 as well as the Cournot triopoly with bounded rationality \cite{ABK1999,AGP2000} has been studied.
All these studies report that the dynamics of an oligopoly with bounded rationality is very complicated.

Another significant concept in modern microeconomics is information asymmetry, where one player has more or better information than the others. 
This asymmetry can create a power imbalance in transactions, which in turn could lead to market failure in the worst case.
Previous studies have considered the effect of information asymmetry on games with rational players \cite{MWG1995}.
However, the concept of information asymmetry in games where the players have bounded rationality has not yet been well established.
Therefore, we try to explain the effect of information asymmetry on games with bounded rationality by taking the Cournot duopoly game as an example.

Concretely, we investigate a discrete-time dynamic Cournot duopoly game of players with bounded rationality where one player has information about the behavior of the other player.
The player with information adjusts his output based on the present output of the other player. 
We then theoretically and numerically show that the information stabilizes the Cournot-Nash equilibrium and suppresses the chaotic behavior in larger parameters compared with the case where there is no information asymmetry.
A similar situation was investigated in the context of information acquisition by two information-sharing firms about the third firm in a triopoly game \cite{GuoMa2016}.

The paper is organized as follows.
Section \ref{sec:model} introduces a dynamic Cournot duopoly game model of bounded rational players with information asymmetry.
Section \ref{sec:results} provides theoretical and numerical results that the Cournot-Nash equilibrium is stabilized and chaotic behavior is suppressed by effect of information asymmetry.
Section \ref{sec:conclusion} provides the concluding remarks of the paper.
\ref{app:original} reviews the results for the dynamic Cournot duopoly game where there is no information asymmetry; see Ref. \cite{BisNai2000}.

\section{Model}
\label{sec:model}
We consider two firms (players) $i=1, 2$ producing the same good.
The production (output) of each firm is described by $q_i$.
We assume that the price of the good is determined by the total supply $Q = q_1+q_2$ through the linear inverse demand function:
\begin{eqnarray}
 p(q_1, q_2) &=& a-bQ
\end{eqnarray}
where $a$ and $b$ are positive constants.
If the total supply is increased, the price of the good becomes lower.
We also assume that the cost of production of firm $i$ is linear, as $cq_i$ with $c>0$.
With these assumptions, the profit of firm $i$ can be given by
\begin{eqnarray}
 \Pi_i(q_1, q_2) &=& \left( a-bQ \right)q_i - cq_i.
\end{eqnarray}
In the right-hand side, the first term represents the sales of the good and the second term represents the cost of the good. 
Thus, if these difference becomes larger, the profit is increased.
When both players are rational, their best responses are given by the profit maximization conditions
\begin{eqnarray}
 0 &=& \frac{\partial \Pi_1}{\partial q_1}(q_1, q_2)
\end{eqnarray}
and
\begin{eqnarray}
 0 &=& \frac{\partial \Pi_2}{\partial q_2}(q_1, q_2),
\end{eqnarray}
to achieve the Cournot-Nash equilibrium
\begin{eqnarray}
 q_1=q_2=\frac{a-c}{3b}.
\end{eqnarray}
We can easily find that those conditions give the maximum profit, respectively.

We consider the dynamics of two bounded rational players \cite{BisNai2000}.
The production of player $i$ at time $t\in \left\{0, 1, \cdots \right\}$ is described as $q_i(t)$.
Each player adjusts his production step by step in order to increase his profit.
We consider the situation in which player $1$ adjusts his production as
\begin{eqnarray}
 q_1(t+1) &=& q_1(t) + \alpha q_1(t) \frac{\partial \Pi_1}{\partial q_1}\left( q_1(t), q_2(t) \right).
\end{eqnarray}
In the right-hand side, the first term means the production in the previous step, and the second term means the effect that player $1$ adjusts his production step by step according to the gradient of his profit.
The positive constant $\alpha$ is called speed of adjustment.
We consider the information asymmetric situation such that player $2$ already knows $q_1(t+1)$ at time $t+1$ in some way and chooses $q_2(t+1)$ as 
\begin{eqnarray}
 q_2(t+1) &=& q_2(t) + \alpha q_2(t) \frac{\partial \Pi_2}{\partial q_2}\left( q_1(t+1), q_2(t) \right).
\end{eqnarray}
Here, we assume that the two players' speed of adjustment is common.
Therefore, the dynamics of the two players is described as
\begin{eqnarray}
 q_1(t+1) &=& q_1(t) + \alpha q_1(t) \left[ a-c-2bq_1(t) - bq_2(t) \right] \label{eq:q1}\\
 q_2(t+1) &=& q_2(t) + \alpha q_2(t) \left[ a-c-2bq_2(t) - bq_1(t+1) \right] \nonumber \\
 &=& q_2(t) + \alpha q_2(t) \left[ a-c-2bq_2(t) - bq_1(t) \right] \nonumber \\
 && \qquad - \alpha^2 b q_1(t)q_2(t)\left[ a-c-2bq_1(t) - bq_2(t) \right].
\label{eq:q2}
\end{eqnarray}
In \ref{app:original}, we provide the results for the model where there is no information asymmetry.

\section{Results}
\label{sec:results}
\subsection{Equilibrium points and local stability}
We first study the equilibrium points of the dynamic game.
The fixed points are given by equations
\begin{eqnarray}
 0 &=& q_1 \left[ a-c-2bq_1 - bq_2 \right] \label{eq:fp_1} \\
 0 &=& q_2 \left[ a-c-2bq_2 - bq_1 \right],
 \label{eq:fp_2}
\end{eqnarray}
which are obtained by considering $q_1(t)=q_1(t+1)=q_1$ and $q_2(t)=q_2(t+1)=q_2$ in (\ref{eq:q1}) and (\ref{eq:q2}), respectively.
Note that this condition is the same as the case where there is no information asymmetry.
We find that there are four fixed points:
\begin{eqnarray}
 E_0 &=& \left( 0, 0 \right) \\
 E_1 &=& \left( \frac{a-c}{2b}, 0 \right) \\
 E_2 &=& \left( 0, \frac{a-c}{2b} \right) \\
 E_* &=& \left( \frac{a-c}{3b}, \frac{a-c}{3b} \right), \label{eq:fp_CN}
\end{eqnarray}
which are obtained by the condition that each factor in (\ref{eq:fp_1}) and (\ref{eq:fp_2}) becomes zero.
The fixed points $E_1$ and $E_2$ correspond to monopolistic fixed points.
We assume that the Cournot-Nash equilibrium $E_*$ exists; that is, 
\begin{eqnarray}
 a-c &>& 0.
 \label{eq:Nashcondition}
\end{eqnarray}

The local stability of each fixed point is characterized by eigenvalues of the Jacobian matrix
\begin{eqnarray}
 J\left( q_1, q_2 \right) &=& \left(
    \begin{array}{cc}
      J_{1,1} & J_{1,2} \\
      J_{2,1} & J_{2,2}
    \end{array}
  \right)
\end{eqnarray}
with 
\begin{eqnarray}
  J_{1,1} &=& 1+\alpha(a-c-4bq_1-bq_2) \\
  J_{1,2} &=& -\alpha b q_1 \\
  J_{2,1} &=& -\alpha b q_2 - \alpha^2b(a-c)q_2+4\alpha^2b^2q_1q_2 + \alpha^2b^2q_2^2 \\
  J_{2,2} &=& 1+\alpha(a-c-4bq_2-bq_1) - \alpha^2b(a-c)q_1+2\alpha^2b^2q_1^2 \nonumber \\
  && \qquad + 2 \alpha^2b^2q_1q_2.
\end{eqnarray}
When the absolute value of the two eigenvalues is smaller than $1$, displacement from the fixed point decays to zero, that is, the fixed point is locally stable.
First, we find that
\begin{eqnarray}
 J\left( 0, 0 \right) &=& \left(
    \begin{array}{cc}
      1+A & 0 \\
      0 &  1+A
    \end{array}
  \right)
\end{eqnarray}
with $A\equiv \alpha(a-c)$, and then we find that $E_0$ is unstable under the condition (\ref{eq:Nashcondition}).

Next, we consider the stability of $E_1$ and $E_2$.
The Jacobian matrix at $E_1$ is
\begin{eqnarray}
 J\left( \frac{a-c}{2b}, 0 \right) &=& \left(
    \begin{array}{cc}
      1-A & -\frac{1}{2}A \\
      0 &  1+\frac{1}{2}A
    \end{array}
  \right).
\end{eqnarray}
Therefore, $E_1$ is a saddle point for $0<A<2$ and unstable for $A>2$.
The Jacobian matrix at $E_2$ is
\begin{eqnarray}
 J\left( 0, \frac{a-c}{2b} \right) &=& \left(
    \begin{array}{cc}
      1+\frac{1}{2}A & 0 \\
      - \frac{1}{2}A - \frac{1}{4}A^2 &  1-A
    \end{array}
  \right).
\end{eqnarray}
We observe that $E_2$ is also a saddle point for $0<A<2$ and unstable for $A>2$.

Finally, we investigate the stability of the Cournot-Nash fixed point $E_*$.
The Jacobian matrix at $E_*$ is
\begin{eqnarray}
 J\left( \frac{a-c}{3b}, \frac{a-c}{3b} \right) &=& \left(
    \begin{array}{cc}
      1-\frac{2}{3}A & -\frac{1}{3}A \\
      - \frac{1}{3}A + \frac{2}{9}A^2 &  1-\frac{2}{3}A+\frac{1}{9}A^2
    \end{array}
  \right).
\end{eqnarray}
The eigenvalues of this matrix are given by
\begin{eqnarray}
 \lambda &=& \frac{1}{18} \left[ 18-12A+A^2 \pm A\sqrt{36-24A+A^2} \right].
\end{eqnarray}
Both eigenvalues are non-negative and $\lambda<1$ for $0<A<12-6\sqrt{3}$.
For $A>12-6\sqrt{3}$, $\lambda$ is complex and 
\begin{eqnarray}
 |\lambda|^2 &=& \left( 1 - \frac{2}{3}A \right)^2.
\end{eqnarray}
Therefore, $E_*$ is locally stable for $A<3$.
According to \ref{app:original}, the stability condition of $E_*$ in the original Cournot game is $A<2$.
Therefore, we can say that the information asymmetry broadens the stabilized region of $E_*$.

\subsection{Bifurcation diagram}
\begin{figure}[t]
\includegraphics[clip, width=8.0cm]{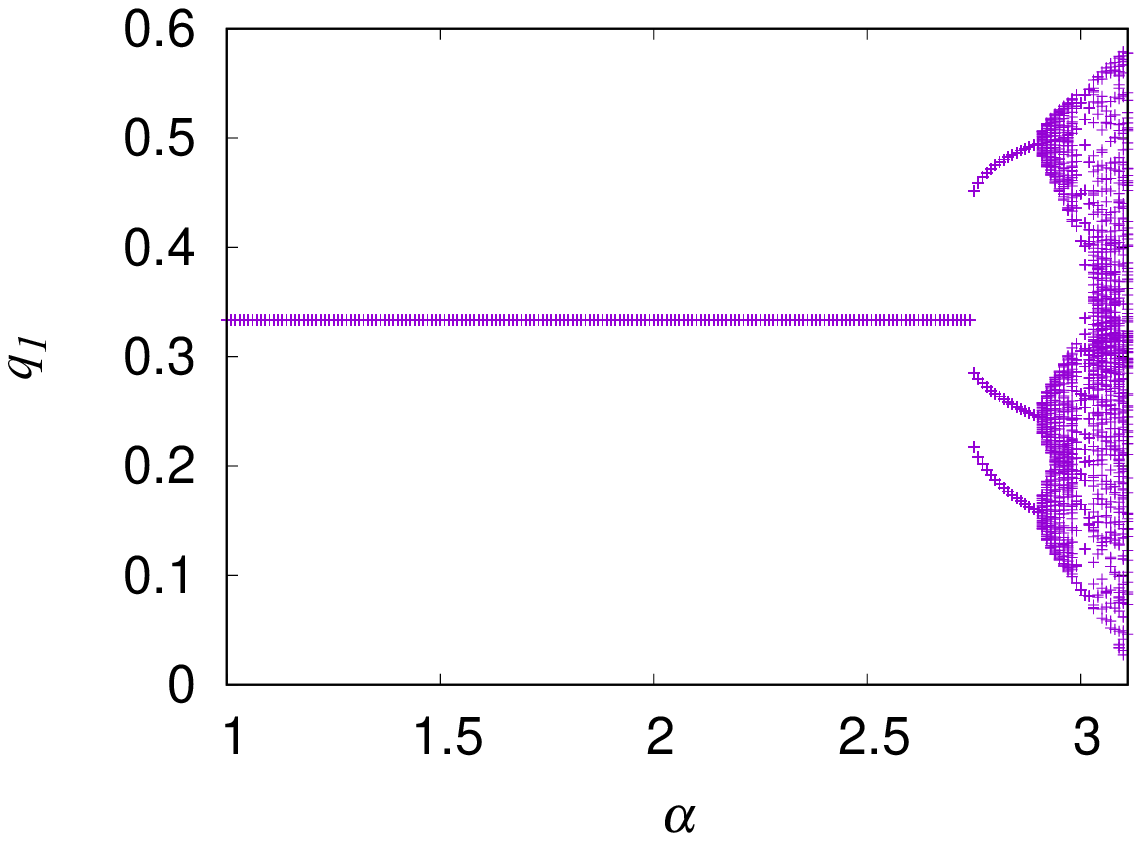}
\includegraphics[clip, width=8.0cm]{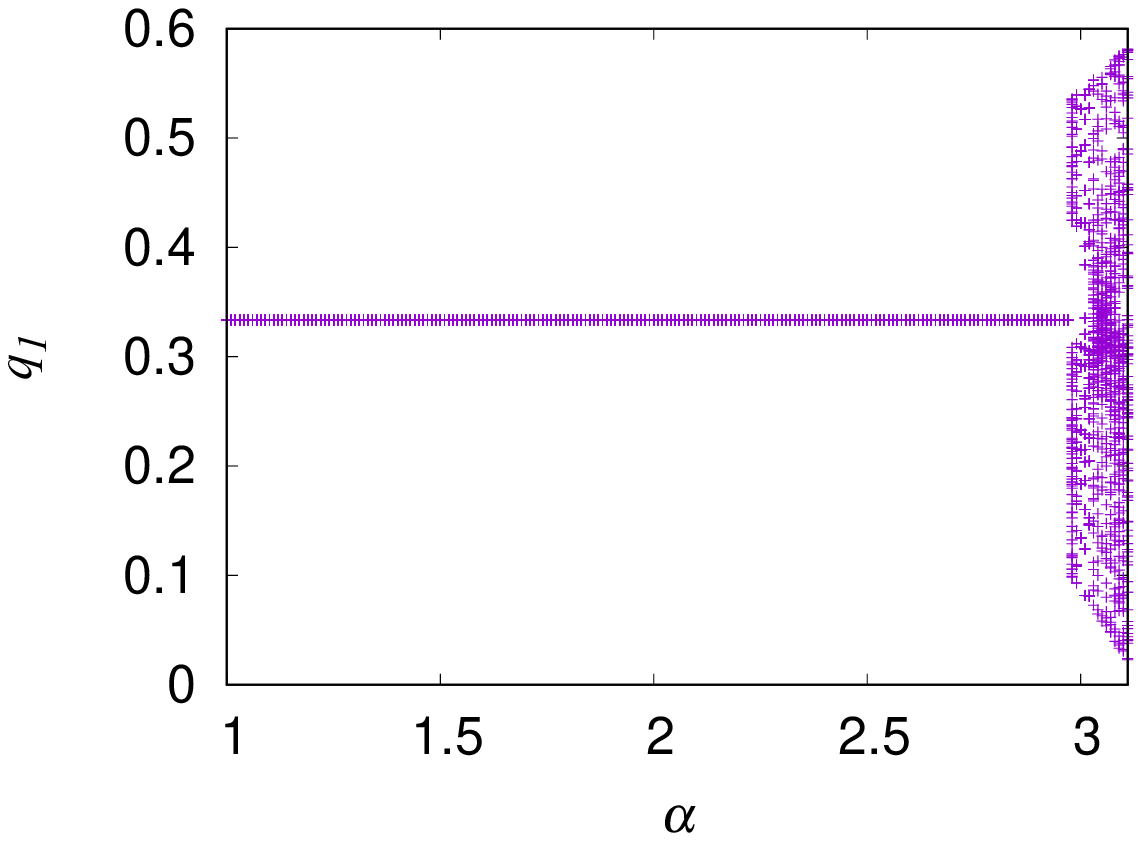}
\caption{Bifurcation diagram for $q_1$ with $a=2$, $b=1$, and $c=1$, and initial condition $\left(q_1(0), q_2(0)\right)=(0.1,0.1)$ (top) and $\left(q_1(0), q_2(0)\right)=(0.33,0.33)$ (bottom). It should be noted that the Cournot-Nash fixed point (\ref{eq:fp_CN}) is $(1/3, 1/3)$.}
\label{fig:bif_infasym}
\end{figure}

In order to investigate properties of trajectories realized by our model, we numerically solve equations (\ref{eq:q1}) and (\ref{eq:q2}).
We set the parameters as $a=2$, $b=1$, and $c=1$.
In these parameters, we have $E_*=(1/3,1/3)$ and $A=\alpha$.
Fig. \ref{fig:bif_infasym} plots the bifurcation diagram of $q_1$ with initial condition $\left(q_1(0), q_2(0)\right)=(0.1,0.1)$, and $\left(q_1(0), q_2(0)\right)=(0.33,0.33)$.
Points $q_1(t)$ with $t\in [1001,1100]$ are plotted for each $\alpha$.
We observe that a periodic orbit with period three discontinuously appears at $\alpha\simeq 2.75$ for the former case, although the local stability condition $\alpha<3$ of the Cournot-Nash fixed point $E_*$ is satisfied in this region.
That is, although the Cournot-Nash equilibrium is locally stable in $\alpha<3$, it is not necessarily globally stable.
Such discontinuous appearance of a periodic trajectory does not occur for an information symmetric case, as seen in \ref{app:original}.
We can also see aperiodic behavior for larger $\alpha$.
Note that the trajectories become unbounded for $\alpha>3.11$.

We can understand the appearance of a periodic orbit as follows.
If equations (\ref{eq:q1}) and (\ref{eq:q2}) have the non-trivial solution $\left( q_1(t), q_2(t) \right)=(r_2, r_1)$, $\left( q_1(t+1), q_2(t+1) \right)=(r_3, r_2)$, and $\left( q_1(t+2), q_2(t+2) \right)=(r_1, r_3)$ with
\begin{eqnarray}
 r_1 &=& r_3 + \alpha r_3 \left[ a - c - 2br_3 - br_2 \right] \\
 r_2 &=& r_1 + \alpha r_1 \left[ a - c - 2br_1 - br_3 \right] \\
 r_3 &=& r_2 + \alpha r_2 \left[ a - c - 2br_2 - br_1 \right],
\end{eqnarray}
a periodic orbit exists.
When we define $R_n\equiv \alpha b r_n$ with $n=1,2,3$, these $R_n$ satisfy 
\begin{eqnarray}
 R_1 &=& R_3 + R_3 \left[ A - 2R_3 - R_2 \right] \\
 R_2 &=& R_1 + R_1 \left[ A - 2R_1 - R_3 \right] \\
 R_3 &=& R_2 + R_2 \left[ A - 2R_2 - R_1 \right].
\end{eqnarray}
These equations have a non-trivial solution for $A$ that is larger than $A\simeq 2.75$.
This explains the existence of a periodic solution.

\begin{figure}[t]
\includegraphics[clip, width=8.0cm]{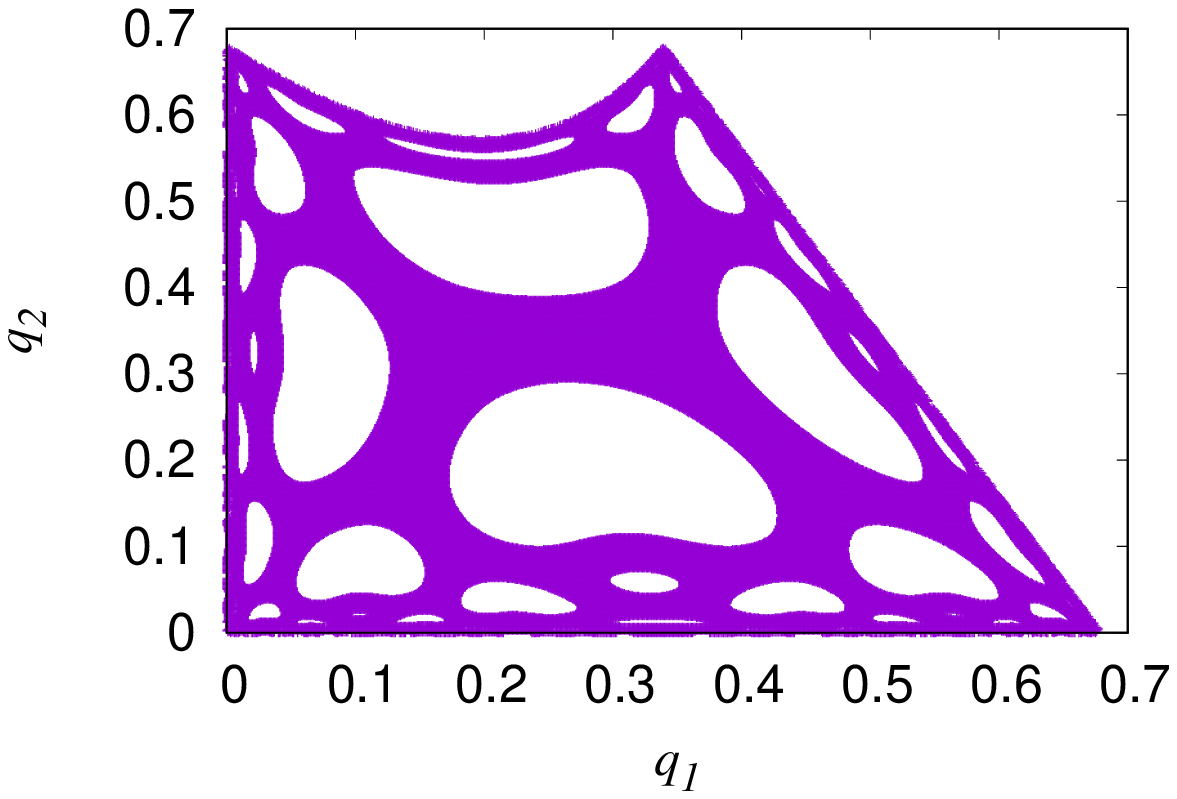}
\includegraphics[clip, width=8.0cm]{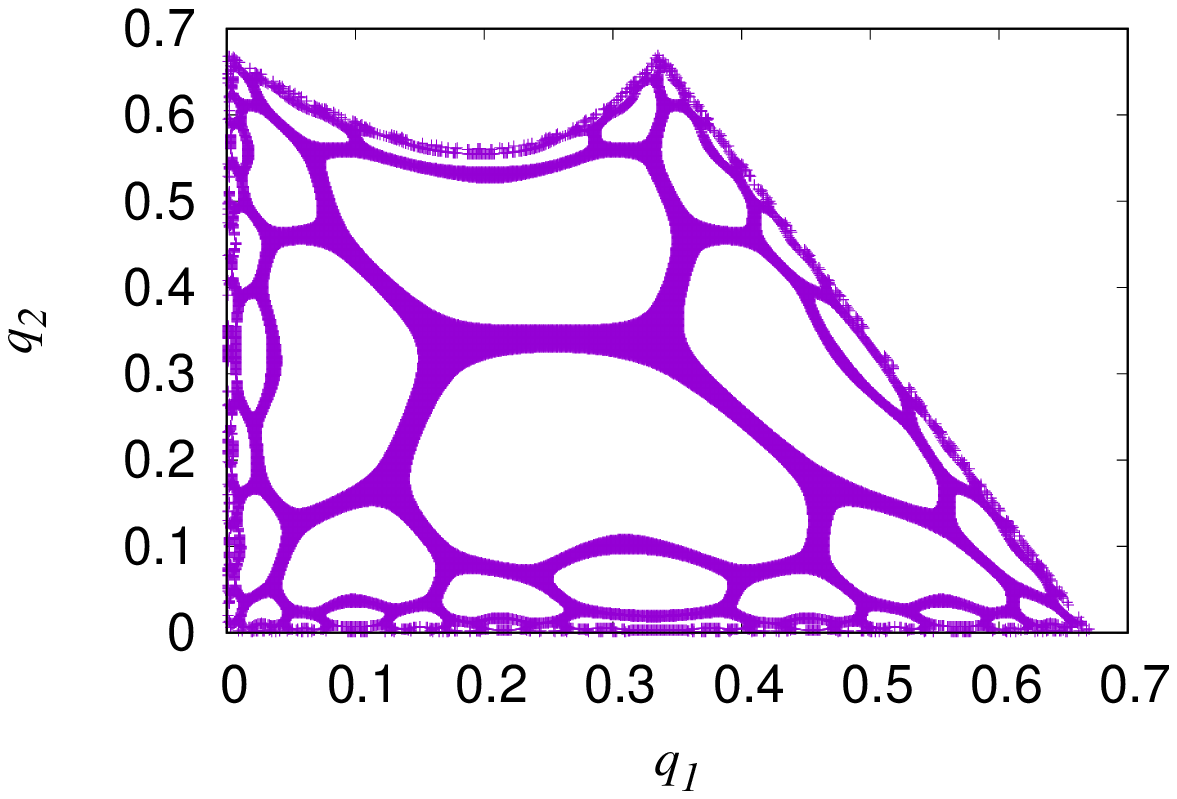}
\caption{The basin of attraction of the Cournot-Nash equilibrium point $E_*$ at $\alpha=2.8$ (top) and $\alpha=2.9$ (bottom).}
\label{fig:basin_infasym}
\end{figure}

We next discuss the basin of attraction of the Cournot-Nash equilibrium point $E_*$.
In Fig. \ref{fig:basin_infasym}, we display the basin of attraction of $E_*$ at $\alpha=2.8$ and $\alpha=2.9$, which is the set of initial conditions $(q_1(0), q_2(0))$ which converge to the Cournot-Nash fixed point $E_*$ after $2000$ iterations.
We observe that the basin of attraction of $E_*$ becomes smaller and smaller as $\alpha$ increases.
It should be noted that the basin of attraction disappears at $\alpha=3.0$.
Therefore, the Cournot-Nash equilibrium point is not globally stable after the periodic orbit appears.

\subsection{Maximal Lyapunov exponent}
The Lyapunov exponent is a quantity in dynamical systems theory that characterizes the rate of separation of infinitesimally close trajectories.
When the Lyapunov exponent is positive, it implies that behavior of the dynamical system is chaotic.
In contrast, when it is negative, the separation of infinitesimally close trajectories converges to zero.
Generally, the Lyapunov exponent depends on the direction of the initial separation vector.
Therefore, the number of the Lyapunov exponents is equal to the dimension of the phase space.
Because the existence of chaotic behavior is characterized by the maximal Lyapunov exponent, which is the largest Lyapunov exponent, we focus on the maximal Lyapunov exponent.
 
\begin{figure}[t]
\includegraphics[clip, width=8.0cm]{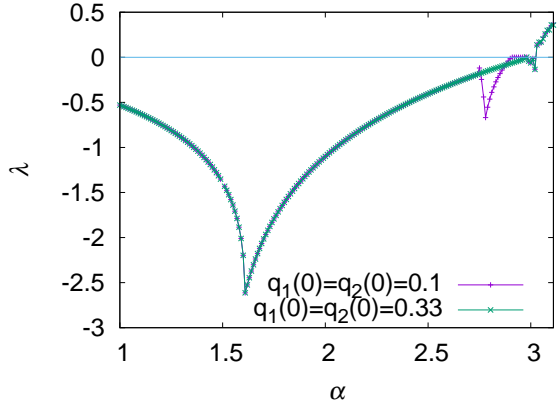}
\caption{Maximal Lyapunov exponent for $a=2$, $b=1$, and $c=1$, and initial condition $\left(q_1(0), q_2(0)\right)=(0.1,0.1)$ and $\left(q_1(0), q_2(0)\right)=(0.33,0.33)$. The straight line represents zero.}
\label{fig:MLE_infasym}
\end{figure}

\begin{figure}[t]
\includegraphics[clip, width=8.0cm]{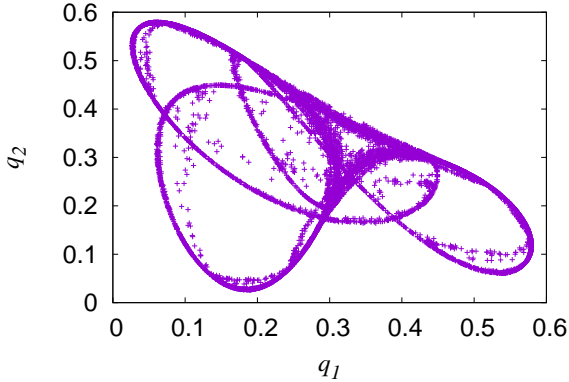}
\caption{A chaotic attractor at $\alpha=3.10$.}
\label{fig:attractor_infasym}
\end{figure}

We numerically calculate the maximal Lyapunov exponent by using the method proposed in Ref. \cite{ShiNag1979}.
Fig. \ref{fig:MLE_infasym} displays the maximal Lyapunov exponent for $a=2$, $b=1$, and $c=1$, and the initial condition $\left(q_1(0), q_2(0)\right)=(0.1,0.1)$ and $\left(q_1(0), q_2(0)\right)=(0.33,0.33)$.
We observe that the maximal Lyapunov exponent is positive for $\alpha>3.02$, implying a chaotic behavior for $\alpha>3.02$.
We also find that the bifurcation from a periodic orbit with period three at $\alpha\sim 2.90$ does not seem to contribute to chaotic behavior.
Fig. \ref{fig:attractor_infasym} displays a chaotic attractor at $\alpha=3.10$.

\section{Concluding remarks}
\label{sec:conclusion}
In this paper, we investigated the effect of information asymmetry on the discrete-time dynamic Cournot duopoly game with bounded rationality.
Concretely, we studied how one player's information about the other player's behavior in a duopoly affects the stability of the Cournot-Nash equilibrium.
We theoretically and numerically showed that the information stabilizes the Cournot-Nash equilibrium and suppresses the chaotic behavior.
Note that a periodic orbit does not discontinuously appear in case there is no information asymmetry.
Our result suggests that information acquisition about the strategy of the other firm in oligopoly market is useful to stabilize the equilibrium point.
This property may hold in realistic market.
The case that speeds of adjustment of two players are not common will be studied in future.

The interpretation of our results from the perspective of numerical simulation suggests that the non-synchronous update of $q_i$ can avoid chaotic behavior in this model.
A future study should examine whether a similar behavior is observed for different games with bounded rationality, which also report chaotic behavior \cite{SAF2002,GalFar2013}.

\section*{Acknowledgement}
This study was supported by JSPS KAKENHI Grant Numbers JP18H06476 and JP19K21542.

\appendix

\section{}
\label{app:original}
In this appendix, we review the results for the case where player $2$ does not have information about the current behavior of player $1$; see Ref. \cite{BisNai2000}.
The dynamics of the two players in this case is described by the map
\begin{eqnarray}
 q_1(t+1) &=& q_1(t) + \alpha q_1(t) \left[ a-c-2bq_1(t) - bq_2(t) \right] \\
 q_2(t+1) &=& q_2(t) + \alpha q_2(t) \left[ a-c-2bq_2(t) - bq_1(t) \right].
\end{eqnarray}
That is, two players are symmetric and bounded rational.
The fixed points are given by the equations
\begin{eqnarray}
 0 &=& q_1 \left[ a-c-2bq_1 - bq_2 \right] \\
 0 &=& q_2 \left[ a-c-2bq_2 - bq_1 \right].
\end{eqnarray}
Here, we find four fixed points:
\begin{eqnarray}
 E_0 &=& \left( 0, 0 \right) \\
 E_1 &=& \left( \frac{a-c}{2b}, 0 \right) \\
 E_2 &=& \left( 0, \frac{a-c}{2b} \right) \\
 E_* &=& \left( \frac{a-c}{3b}, \frac{a-c}{3b} \right).
\end{eqnarray}
We assume the Cournot-Nash equilibrium $E_*$ exists; that is, 
\begin{eqnarray}
 a-c &>& 0.
 \label{eq:Nashcondition_original}
\end{eqnarray}

The stability of each fixed point is characterized by eigenvalues of the Jacobian matrix
\begin{eqnarray}
 J\left( q_1, q_2 \right) &=& \left(
    \begin{array}{cc}
      J_{1,1} & J_{1,2} \\
      J_{2,1} &  J_{2,2}
    \end{array}
  \right)
\end{eqnarray}
with 
\begin{eqnarray}
  J_{1,1} &=& 1+\alpha(a-c-4bq_1-bq_2) \\
  J_{1,2} &=& -\alpha b q_1 \\
  J_{2,1} &=& -\alpha b q_2 \\
  J_{2,2} &=& 1+\alpha(a-c-4bq_2-bq_1).
\end{eqnarray}
First, we find that
\begin{eqnarray}
 J\left( 0, 0 \right) &=& \left(
    \begin{array}{cc}
      1+A & 0 \\
      0 &  1+A
    \end{array}
  \right)
\end{eqnarray}
with $A\equiv \alpha(a-c)$, and then we find that $E_0$ is unstable under the condition (\ref{eq:Nashcondition_original}).

Next, we consider the stability of $E_1$.
The Jacobian matrix at $E_1$ is
\begin{eqnarray}
 J\left( \frac{a-c}{2b}, 0 \right) &=& \left(
    \begin{array}{cc}
      1-A & -\frac{1}{2}A \\
      0 &  1+\frac{1}{2}A
    \end{array}
  \right).
\end{eqnarray}
Therefore, $E_1$ is the saddle point for $0<A<2$ and unstable for $A>2$.
Because the two players are symmetric, this result also holds for $E_2$.

Finally, we investigate the stability of the Cournot-Nash fixed point $E_*$.
The Jacobian matrix at $E_*$ is
\begin{eqnarray}
 J\left( \frac{a-c}{3b}, \frac{a-c}{3b} \right) &=& \left(
    \begin{array}{cc}
      1-\frac{2}{3}A & -\frac{1}{3}A \\
      - \frac{1}{3}A &  1-\frac{2}{3}A
    \end{array}
  \right).
\end{eqnarray}
The eigenvalues of this matrix are
\begin{eqnarray}
 \lambda &=& 1-A, \qquad 1-\frac{1}{3}A.
\end{eqnarray}
Therefore, the local stability condition of $E_*$ is $A < 2$.

\begin{figure}[t]
\includegraphics[clip, width=8.0cm]{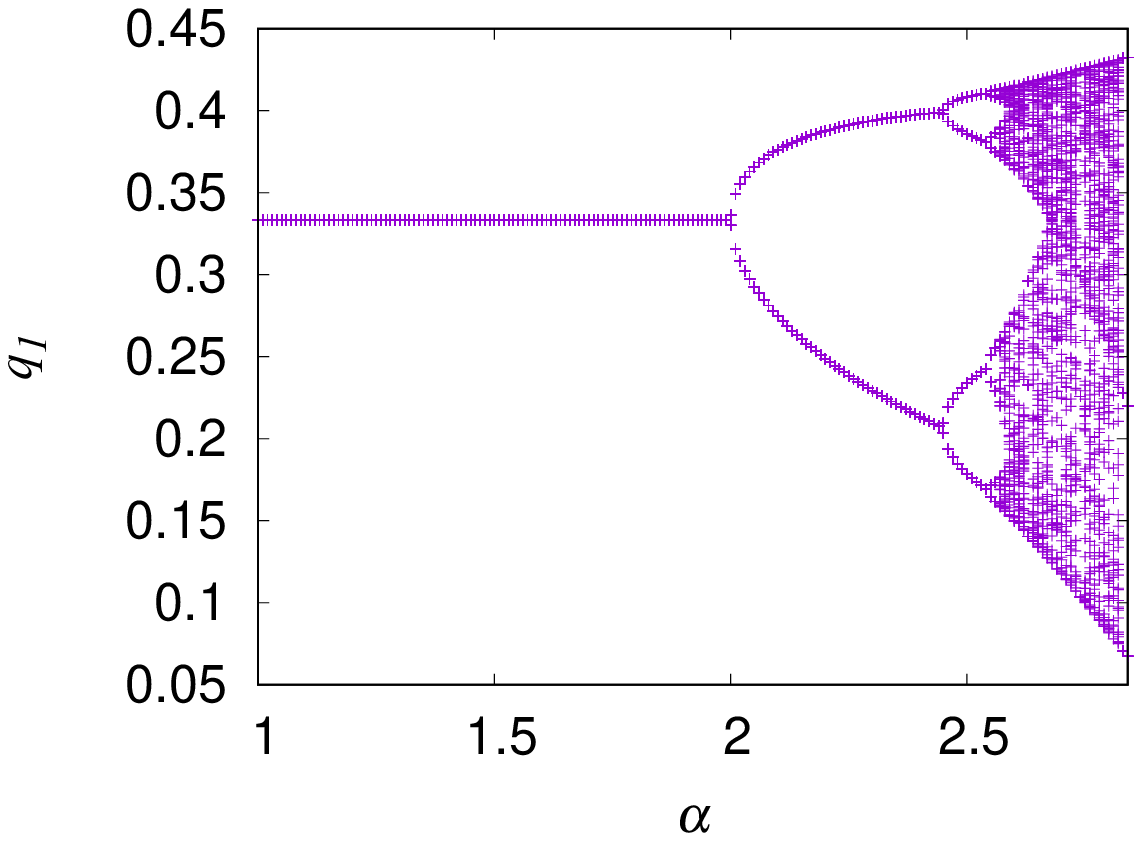}
\caption{Bifurcation diagram for $q_1$ with $a=2$, $b=1$, and $c=1$, with initial condition $\left(q_1(0), q_2(0)\right)=(0.1,0.2)$.}
\label{fig:bif_original}
\end{figure}

\begin{figure}[t]
\includegraphics[clip, width=8.0cm]{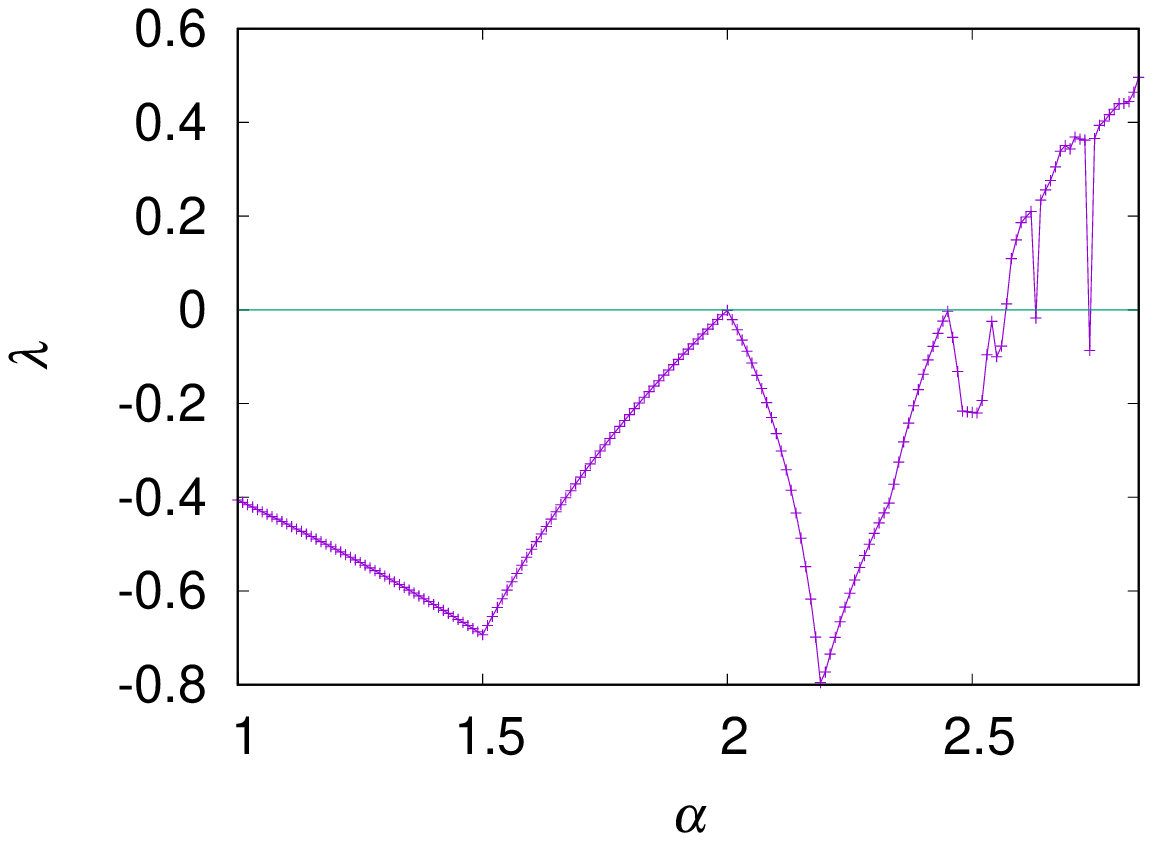}
\caption{Maximal Lyapunov exponent for $a = 2$, $b = 1$, and $c = 1$, and initial condition $\left(q_1(0), q_2(0)\right) = (0.1,0.2)$. The straight line represents zero.}
\label{fig:MLE_original}
\end{figure}

We present the bifurcation diagram in Fig. \ref{fig:bif_original}.
The parameters are set to $a = 2$, $b = 1$, and $c=1$.
The initial condition is set to $\left(q_1(0), q_2(0)\right)=(0.1,0.2)$, and $100$ points are plotted after $1000$ iterations.
We find period-doubling bifurcation occurring at $\alpha = 2$.
Note that the trajectories become unbounded for $\alpha>2.84$.
We also display the $\alpha$ dependence of maximal Lyapunov exponent in Fig. \ref{fig:MLE_original}.
We observe that the maximal Lyapunov exponent is positive for $\alpha$ larger than $\alpha_*\simeq 2.57$, indicating chaotic behavior.

\section*{References}

\bibliography{cournot_bib}

\end{document}